# Alternative Spreadsheet Model Designs for an Operations Management Model Embedded in a Periodic Business Process

Thomas A. Grossman
Vijay Mehrotra
Mouwafac Sidaoui

University of San Francisco
2130 Fulton Street, San Francisco, CA 94117-1045, USA

tagrossman@usfca.edu, vmehrotra@usfca.edu, sidaoui@usfca.edu

**ABSTRACT**

We present a widely-used operations management model used in supply and distribution planning, that is typically embedded in a periodic business process that necessitates model modification and reuse. We consider three alternative spreadsheet implementations, a data-driven design, a canonical (textbook) design, and a novel (table-driven) technical design. We evaluate each regarding suitability for accuracy, modification, analysis, and transfer. We consider the degree of training and technical sophistication required to utilize each design. The data-driven design provides insight into poor spreadsheet practices by naïve modelers. The technical design can be modified for new data and new structural elements without manual writing or editing of cell formulas, thus speeding modification and reducing risk of error. The technical design has potential for use with other classes of models. We identify opportunities for future research.

## 1. INTRODUCTION

There has been much attention in the literature to the design of spreadsheet models for financial analysis. There are several methodologies (including SSRB 2016, FAST 2016,and Operis2017) that are compared in [Grossman and Özlük2010].However, relatively little attention has been given to operational planning and analysis models, another area in which spreadsheets are widely used. This paper examines an operational model that has received scant attention in the spreadsheet literature despite being well-known in practice and in the operations management and operations research fields.

### 1.1 Overview

We describe the model in section 2. In section 3, we present three different designs to implement the model in a spreadsheet: a "data-driven" design, a "canonical" or textbook design, and a novel "technical" design that utilizes Excel Tables. We evaluate these designs according to a set of quality criteria, and also discuss the level of spreadsheet training and sophistication necessary for an operations analyst or manager to utilize each design. In section 4, we summarize the three designs and compare their suitability according to the quality criteria. We conclude and provide suggestions for further research in section 5.

### 1.2 Contribution

We use a model that is well-known in the operations management and operations research literature as our foundation. This paper presents different spreadsheet implementations of a single model and explores their provenance. We evaluate each on a set of quality criteria, and level of sophistication and training needed to effectively utilize it. We introduce a spreadsheet design which we see frequently with naïve modelers. This helps us to understand the root cause of certain poor spreadsheet practices that are known to be problematic. We diagnose the problem as a "data-driven" spreadsheet design. We contrast the data-driven design with the well-known "canonical" design from the textbooks. We also present a new "technical" design that takes advantage of Excel's Table feature to implement this model in a manner that makes even extensive modification very easy, *without the need to write or edit any cell formulas whatsoever*. This technical design has potential to be applicable to other classes of model, and we compare it to the data-driven and canonical approaches.

## 2. ANALYTIC SPREADSHEET MODELS FOR DISTRIBUTION AND SOURCING

We present a model that is representative of a class of distribution and sourcing problems. This model serves as the foundation for the rest of the paper.

### 2.1 Overview

We consider an analytic spreadsheet model for distribution and sourcing. In this model, a number of suppliers provide a product (such as raw materials, parts, or finished goods) to a number of destinations(such as factories, warehouses, or markets) that have a demand for it. Each destination must receive a specified quantity of the product ("requirements"). Each supplier has a maximum available quantity of the product ("capacity"). There is a cost associated with each supplier-destination combination.

The operations manager for the company that manages the destinations is interested in "costing out" different plans for sourcing the product from the suppliers. In assessing these sourcing plans, she may also be interested in evaluating other risk characteristics, such as excess supplier capacity available to meet increased demand levels.

### 2.2 Example: Canonical Design

In Figure 1, we present a specific example of this type of distribution and sourcing model, using the "canonical" design that is discussed in more detail in section 3.2. In this example, there are three destinations (listed in columns C-E and in columns J-L) and ten suppliers (listed in rows 6 to 15).

| | A | B | C | D | E | F | G | H | I | J | K | L | M | N | O |
|---|---|---|---|---|---|---|---|---|---|---|---|---|---|---|---|
| 1 | | | | | | | | | | | | | | | |
| 2 | | | | STRUCTURED DATA | | | | | | | SOURCING PLAN AND MODEL | | | | |
| 3 | | | | | | | | | | | | | | | |
| 4 | | | | Costs ($/unit) | | | | | | | Sourcing Plan (Units Shipped) | | | | |
| 5 | | | Abbot | Bone | Chest | | Capacity | | | Abbot | Bone | Chest | | Supplied | |
| 6 | | Georgican | | | $ 30.80 | | 2,500 | | Georgican | | | 1,000 | | 1,000 | |
| 7 | | Hickock | | | $ 36.80 | | 9,000 | | Hickock | | | 1,000 | | 1,000 | |
| 8 | | India | | | $ 34.00 | | 3,000 | | India | | | 1,000 | | 1,000 | |
| 9 | | Johnson | $ 42.00 | $ 41.60 | $ 45.60 | | 27,000 | | Johnson | 1,000 | 1,000 | 1,000 | | 3,000 | |
| 10 | | Lincoln | $ 33.15 | | | | -6,000 | | Lincoln | 1,000 | | | | 1,000 | |
| 11 | | Manister | $ 32.00 | | | | 3,000 | | Manister | 1,000 | | | | 1,000 | |
| 12 | | Ocean | $ 44.10 | $ 45.36 | | | 30,000 | | Ocean | 1,000 | 1,000 | | | 2,000 | |
| 13 | | Calais | | $ 35.00 | | | 3,600 | | Calais | | 1,000 | | | 1,000 | |
| 14 | | Robert | | $ 33.12 | | | 2,700 | | Robert | | 1,000 | | | 1,000 | |
| 15 | | Simpson | | $ 32.40 | | | 2,300 | | Simpson | | 1,000 | | | 1,000 | |
| 16 | | | | | | | | | | | | | | | |
| 17 | | Required | 32,422 | 21,233 | 25,125 | | | | Delivered | 4,000 | 5,000 | 4,000 | | | |
| 18 | | | | | | | | | | | | | | | |
| 19 | | | | | | | | | | Total Sourcing Cost | $ | 485,930 | | | |
| 20 | | | | | | | | | | | | | | | |

*Figure 1:Canonical Design for a Simple Distribution and Sourcing Model*

The model inputs are the cost coefficients (C6:E15), the supplier capacity levels (G6:G15), the requirements for each destination (C17:E17),and the sourcing plan (J6:L15). The model outputs are the total sourcing cost (K19), total supplied by each supplier (N6:N15), and total quantity delivered to each destination (J17:L17).

The cells that are grayed out correspond to supplier-destination combinations that are not available (due to reasons of quality, prohibitive cost, or other issues).The particular sourcing plan shown in Figure 1 has placeholder values of 1000 units of product from each supplier to each allowable destination.

The operations manager will experiment with different values for the sourcing plan in order to obtain insight into the sourcing plan she will ultimately commit to. This sourcing plan (J6:L15) is of great practical importance because it determines the quantity shipped from each supplier to each destination, the total amount supplied by each supplier, the total amount delivered to each destination, the total amount paid to each supplier, and whether a supplier is even selected to do business with the company.

We note that the model in Figure 1 represents a single time period. In the general case, the time period may be anywhere from a single day to an entire year. Of course, this business operates over time, and so in the next time period a new sourcing plan will be needed. Naturally, future time periods can have different requirements at the destinations, different capacities for the suppliers, or even different suppliers or destinations. Therefore, the model needs to be easy to modify so that it can be employed as part of a routine periodic business process.

### 2.3 Transportation Models Are Commonly Found in Spreadsheets

This model is an instance of what the operations management and operations research communities call the "transportation model" where "sources" (e.g., suppliers) provide resources to "sinks" (e.g., destinations) and each source-sink combination has an associated cost. Transportation models are applied to problems in manufacturing, logistics and distribution, personnel scheduling, vehicle routing, and inventory planning (see Ahuja et al., 1989 for additional applications).

Spreadsheets have proven to be very valuable tools for building and analyzing transportation models(LeBlanc et al. 2004, Shaoyun et al. 2011, Hong et al. 2012).The standard business school textbooks on spreadsheet modeling and analysis discuss transportation models, including [Ragsdale, 2012], [Winston and Albright, 2012], [Powell and Baker, 2013], [Gross et al., 2013], and [Asllani, 2014].For the most part, textbook authors use the model as a basis for algorithms to identify an optimal plan, and pay minimal attention to issues of spreadsheet design.

## 3. SPREADSHEET DESIGNS FOR TRANSPORTATION MODELS

We present three designs that implement a realistic instance of a transportation model. We evaluate each design according to a set of quality criteria drawn from [Grossman et al 2011] including suitability for:

- Accuracy (of formulas; ease of coding)
- Modification (efficiency and accuracy) to enable reuse
- Analysis (use of model to generate business insight)
- Transfer (ease of understanding by other personnel)

We also discuss the level of spreadsheet training and technical sophistication necessary for an operations analyst or manager to utilize each design.

### 3.1 Data-Driven Design

We provide a (somewhat more elaborate) version of this type of sourcing and distribution problem to business school students every year in our courses at the University of San Francisco. The students' assignment is to develop a high-quality analytic spreadsheet model, and then use their model to conduct analysis and deliver recommendations to management.

The students are given this problem in the form of a case study that includes a set of data. Figure 2 shows the data that is provided. As is typical in business, (but not in textbooks), the data is provided in a form convenient for the person (or system) that generated it, but it may or may not be convenient for development of a spreadsheet model.

| A | B | C | D | E |
|---|---|---|---|---|
| | RAW DATA | | | |
| | | | | |
| | **Supplier** | **Destination** | **Shipping Cost/Unit ($)** | **Supplier Capacity** |
| | Georgican | Chest | $ 30.80 | 2,500 |
| | Hickock | Chest | $ 36.80 | 9,000 |
| | India | Chest | $ 34.00 | 3,000 |
| | Johnson | Abbot | $ 42.00 | 27,000 |
| | | Bone | $ 41.60 | |
| | | Chest | $ 45.60 | |
| | Lincoln | Abbot | $ 33.15 | 6,000 |
| | Manister | Abbot | $ 32.00 | 3,000 |
| | Ocean | Abbot | $ 44.10 | 30,000 |
| | | Bone | $ 45.36 | |
| | Calais | Bone | $ 35.00 | 3,600 |
| | Robert | Bone | $ 33.12 | 2,700 |
| | Simpson | Bone | $ 32.40 | 2,300 |
| | | | | |
| | **Total Required By Destination** | | | |
| | Abbot | 32,422 | | |
| | Bone | 21,233 | | |
| | Chest | 25,125 | | |

*Figure 2:Raw Data for Transportation Example*

Inevitably, a minority of our students scorn their professors' guidance to pause and purposefully create a sound spreadsheet design. Instead, they implement the model by immediately writing cell formulas on the same worksheet as the raw data in Figure 2. Thus arises what we call a "Data-driven Design", shown in Figures 3 and 4.

RAW DATA | SOURCING PLAN

| Supplier | Destination | Shipping Cost/Unit ($) | Supplier Capacity | Units Shipped |
|---|---|---|---|---|
| Georgican | Chest | $ 30.80 | 2,500 | 1000 |
| Hickock | Chest | $ 36.80 | 9,000 | 1000 |
| India | Chest | $ 34.00 | 3,000 | 1000 |
| Johnson | Abbot | $ 42.00 | 27,000 | 1000 |
| | Bone | $ 41.60 | | 1000 |
| | Chest | $ 45.60 | | 1000 |
| Lincoln | Abbot | $ 33.15 | 6,000 | 1000 |
| Manister | Abbot | $ 32.00 | 3,000 | 1000 |
| Ocean | Abbot | $ 44.10 | 30,000 | 1000 |
| | Bone | $ 45.36 | | 1000 |
| Calais | Bone | $ 35.00 | 3,600 | 1000 |
| Robert | Bone | $ 33.12 | 2,700 | 1000 |
| Simpson | Bone | $ 32.40 | 2,300 | 1000 |

| Total Required By Destination | |
|---|---|
| Abbot | 32,422 |
| Bone | 21,233 |
| Chest | 25,125 |

MODEL CALCULATIONS

| Total Shipped By Supplier | |
|---|---|
| Georgican | 1000 |
| Hickock | 1000 |
| India | 1000 |
| Johnson | 3000 |
| Lincoln | 1000 |
| Manister | 1000 |
| Ocean | 2000 |
| Calais | 1000 |
| Robert | 1000 |
| Simpson | 1000 |

| Total Sourcing Cost | $ 485,930 |
|---|---|

| Destination | Units Delivered | Units Required |
|---|---|---|
| Abbot | 4000 | 32,422 |
| Bone | 5000 | 21,233 |
| Chest | 4000 | 25,125 |

*Figure 3:Data-driven Design*

MODEL CALCULATIONS

| Total Shipped By Supplier | |
|---|---|
| Georgican | =F4 |
| Hickock | =F5 |
| India | =F6 |
| Johnson | =SUM(F7:F9) |
| Lincoln | =F10 |
| Manister | =F11 |
| Ocean | =SUM(F12:F13) |
| Calais | =F14 |
| Robert | =F15 |
| Simpson | =F16 |

| Total Sourcing Cost | =SUMPRODUCT(D4:D16,F4:F16) |
|---|---|

| Destination | Units Delivered | Units Required |
|---|---|---|
| Abbot | =F7+F10+F11+F12 | =C19 |
| Bone | =F8+F13+F14+F15+F16 | =C20 |
| Chest | =F4+F5+F6+F9 | =C21 |

*Figure 4:Cell Formulas for Data-Driven Design*

Regarding *accuracy,* this design has obvious risks of error during coding and is inefficient to code because the design does not allow a formula to be written once and copied. To see this in Figure 4, compare the formulas in cells I6 and I7, and in cells L4, L5 and L6.

*Modification* of a data-driven design is problematic. If Hickock begins to ship to Bone, this will lead to an additional row in the raw data below row 5.From here, the formulas for Total Shipped by Hickock (I5) and for Total Delivered to Bone (L5) must be manually adjusted. Adding additional suppliers or destinations would require more extensive changes, taking up more time and increasing the risk of errors. *Analysis* is subtly challenging, because when the analyst experiments with plan values in column F, it is not obvious why it impacts the totals in columns I and L. *Transfer* of the model to others is difficult because the design cannot be understood without careful examination of cell formulas. A data-driven design requires no training or sophistication on the part of the modeler; indeed, it is typically our weakest students who deploy this type of design.

### 3.2Canonical Design

The canonical design uses a matrix to represent the units shipped (see Figures 1 and 5) and is presented universally in the textbooks referenced above. The matrix structure of this design intuitively displays the relationships between suppliers and destinations. Regarding *accuracy*, the coding is efficient and low-risk, because the model outputs of supplied (N6:N15) and delivered (J17:L17) are simple row and column sums, and the formulas can be written once and copied.

| | I | J | K | L | M | N | O |
|---|---|---|---|---|---|---|---|
| 1 | | | | | | | |
| 2 | | | SOURCING PLAN AND MODEL | | | | |
| 3 | | | | | | | |
| 4 | | | Sourcing Plan (Units Shipped) | | | | |
| 5 | | Abbot | Bone | Chest | | Supplied | |
| 6 | Georgican | | | 1000 | | =SUM(J6:L6) | |
| 7 | Hickock | | | 1000 | | =SUM(J7:L7) | |
| 8 | India | | | 1000 | | =SUM(J8:L8) | |
| 9 | Johnson | 1000 | 1000 | 1000 | | =SUM(J9:L9) | |
| 10 | Lincoln | 1000 | | | | =SUM(J10:L10) | |
| 11 | Manister | 1000 | | | | =SUM(J11:L11) | |
| 12 | Ocean | 1000 | 1000 | | | =SUM(J12:L12) | |
| 13 | Calais | | 1000 | | | =SUM(J13:L13) | |
| 14 | Robert | | 1000 | | | =SUM(J14:L14) | |
| 15 | Simpson | | 1000 | | | =SUM(J15:L15) | |
| 16 | | | | | | | |
| 17 | Delivered | =SUM(J6:J15) | =SUM(K6:K15) | =SUM(L6:L15) | | | |
| 18 | | | | | | | |
| 19 | | Total Sourcing Cost | =SUMPRODUCT(C6:E15,J6:L15) | | | | |
| 20 | | | | | | | |

*Figure 5:Cell Formulas for Canonical Design*

This approach requires that the ill-formatted raw data be converted to the matrix design. This requires manual coding of cell formulas. However, the cell formulas (Figure 6) are minimal "pointers" without any mathematical functions, and so their accuracy is easy to verify.

| | A | B | C | D | E | F | G | H |
|---|---|---|---|---|---|---|---|---|
| 1 | | | | | | | | |
| 2 | | | | STRUCTURED DATA | | | | |
| 3 | | | | | | | | |
| 4 | | $/Unit | | Cost per unit | | | | |
| 5 | | | Abbot | Bone | Chest | | Capacity | |
| 6 | | Georgican | | | =Q6 | | =R6 | |
| 7 | | Hickock | | | =Q7 | | =R7 | |
| 8 | | India | | | =Q8 | | =R8 | |
| 9 | | Johnson | =Q9 | =Q10 | =Q11 | | =R9 | |
| 10 | | Lincoln | =Q12 | | | | =-R12 | |
| 11 | | Manister | =Q13 | | | | =R13 | |
| 12 | | Ocean | =Q14 | =Q15 | | | =R14 | |
| 13 | | Calais | | =Q16 | | | =R16 | |
| 14 | | Robert | | =Q17 | | | =R17 | |
| 15 | | Simpson | | =Q18 | | | =R18 | |
| 16 | | | | | | | | |
| 17 | | Required | 32422 | 21233 | 25125 | | | |
| 18 | | | | | | | | |

| | O | P | Q | R |
|---|---|---|---|---|
| 1 | | | | |
| 2 | | RAW DATA | | |
| 3 | | | | |
| 4 | | | | |
| 5 | Supplier | Destination | Shipping Cost/Unit ($) | Supplier Capacity |
| 6 | Georgican | Chest | 30.8 | 2500 |
| 7 | Hickock | Chest | 36.8 | 9000 |
| 8 | India | Chest | 34 | 3000 |
| 9 | Johnson | Abbot | 42 | 27000 |
| 10 | | Bone | 41.6 | |
| 11 | | Chest | 45.6 | |
| 12 | Lincoln | Abbot | 33.15 | 6000 |
| 13 | Manister | Abbot | 32 | 3000 |
| 14 | Ocean | Abbot | 44.1 | 30000 |
| 15 | | Bone | 45.36 | |
| 16 | Calais | Bone | 35 | 3600 |
| 17 | Robert | Bone | 33.12 | 2700 |
| 18 | Simpson | Bone | 32.4 | 2300 |

*Figure 6:Pointer formulas to structure the data for the Canonical Design*

Depending on the structure of data for the new month, *modification* can be easy or challenging. If the supplier-destination combinations are the same, new raw data values can be pasted into the raw data module in columns Q and R of Figure 5, and model cell formulas do not need to be adjusted. If an existing supplier begins to ship to an existing destination that was previously prohibited, the structure of the raw data will be altered, requiring relatively simple modification to the structured data module by updating cell references to the new raw data module. All model calculations will be unchanged.

However, if new suppliers and/or destinations are added, the process for updating the model is cumbersome. For example, to add a new supplier to the model, the analyst must go through the following steps:

1. Insert a new row to the cost coefficient and sourcing plan matrix for the new supplier
2. Update the calculations in the “Delivered” row for each destination
3. Import the new version of the raw data
4. Update the links to the raw data in the cost coefficient matrix and to the Capacity column for existing suppliers and for the new supplier

The steps associated with adding a new destination are analogous to these. While these processes are straightforward, the work associated with them takes time and requires coding effort that is not risk-free.

The canonical design is very convenient for *analysis*. As the analyst experiments with different values of units shipped, the impact on Delivered and Supplied are respectively presented intuitively in the same row and column. The canonical design is easy to *transfer* because the row and column sums are intuitive. A modest amount of training and sophistication is required to discover (or know) the matrix design and how to code it.

### 3.3 Technical Design

In this section, we present a “technical” design that codes the model at a higher level of abstraction, leveraging the power of Excel’s Table feature. This allows for more efficient coding and dramatically more efficient model modification. The presentation assumes that the reader is familiar with the function of the Excel Table feature.

The technical design (Figure 7)comprises four Excel tables. The table labeled “Shipping Costs and Sourcing Plan” in B4:E17(designated ‘Ship’internal to Excel’s Table feature) contains information specific to each supplier-destination pair, which is shipping cost data, and units shipped (which is the sourcing plan). The table labeled “Supplier Table” in G3:I14 (‘SupSummary’ for the Table feature)contains information specific to each supplier, which is capacity data, as well as a formula that computes total Supplied. The Table labeled “Destination Table” in K4:M7 (‘DestSummary’ for the Table feature)contains information specific to each destination, which is data on requirements, as well as a formula that computes total delivered. The table labeled “Cost Table” in O4:O5 (‘CostCalc’ for the Table feature) contains information that is not specific to suppliers and destinations, which is the total shipping cost.

**Shipping Costs and Sourcing Plan**
**Table name - Ship**

| Supplier | Destinatio | Shipping Cost/ Unit ($) | Units Shipped |
|---|---|---|---|
| Georgican | Chest | $ 30.80 | 1000 |
| Hickock | Chest | $ 36.80 | 1000 |
| India | Chest | $ 34.00 | 1000 |
| Johnson | Abbot | $ 42.00 | 1000 |
| Johnson | Bone | $ 41.60 | 1000 |
| Johnson | Chest | $ 45.60 | 1000 |
| Lincoln | Abbot | $ 33.15 | 1000 |
| Manister | Abbot | $ 32.00 | 1000 |
| Ocean | Abbot | $ 44.10 | 1000 |
| Ocean | Bone | $ 45.36 | 1000 |
| Calais | Bone | $ 35.00 | 1000 |
| Robert | Bone | $ 33.12 | 1000 |
| Simpson | Bone | $ 32.40 | 1000 |

**Supplier Table**
**Table Name - SupSummary**

| Supplier | Supplier Capacit | Supplied |
|---|---|---|
| Georgican | 3,600 | 1000 |
| Hickock | 2,500 | 1000 |
| India | 9,000 | 1000 |
| Johnson | 3,000 | 3000 |
| Lincoln | 27,000 | 1000 |
| Manister | 6,000 | 1000 |
| Ocean | 3,000 | 2000 |
| Calais | 30,000 | 1000 |
| Robert | 2,700 | 1000 |
| Simpson | 2,300 | 1000 |
| Paulucci | 15,000 | 3000 |

**Destination Table**
**Table Name - DestSummary**

| Destinatio | Required | Delivered |
|---|---|---|
| Abbot | 32,422 | 5000 |
| Bone | 21,233 | 6000 |
| Chest | 25,125 | 4000 |
| Duluth | 12,555 | 1000 |

**Cost Table**

| Total Sourcing Costs |
|---|
| $ 485,930 |

*Figure 7:Technical Design*

In Figure 7, blue cells are input data. Green cells are the sourcing plan. Orange cells contain cell formulas.

This approach requires that the ill-formatted raw data be converted to the table design. This requires manual coding of cell formulas in the blue input data cells. We note that the supplier-destination pairs in the raw data (Figure 2) are the same as the first three columns of the “Ship” Table, so the data in the table can be populated trivially with minimal pointer formulas similar to what was done in Figure 6. Likewise, the Supplier Capacity values in the “SupSummary” Table and the Required values for each of the Destinations are obtained using minimal pointer formulas.

To calculate the total amount shipped by supplier Georgican in cell I5, we need to sum up column E for those rows where column B is equal to “Georgican”. This calculation is performed using SUMIF (cell I5 of Figure 8).It is coded using a “table-style” formula that uses ‘@Supplier’ to indicate the value in the same row (row 5) Supplier column in the Table (i.e., “Georgican” in cell G5). It uses ‘Ship[Supplier]’ to indicate the column labelled Supplier in Table Ship (column B), and uses ‘Ship[Units Shipped]’ to indicate the column labelled Units Shipped in Table Ship (column E).

**Supplier Table**

| Supplier | Supplier Capacity | Supplied |
|---|---|---|
| Georgican | 3600 | =SUMIF([@Supplier],Ship[Supplier],Ship[Units Shipped]) |
| Hickock | 2500 | =SUMIF([@Supplier],Ship[Supplier],Ship[Units Shipped]) |
| India | 9000 | =SUMIF([@Supplier],Ship[Supplier],Ship[Units Shipped]) |
| Johnson | 3000 | =SUMIF([@Supplier],Ship[Supplier],Ship[Units Shipped]) |
| Lincoln | 27000 | =SUMIF([@Supplier],Ship[Supplier],Ship[Units Shipped]) |
| Manister | 6000 | =SUMIF([@Supplier],Ship[Supplier],Ship[Units Shipped]) |
| Ocean | 3000 | =SUMIF([@Supplier],Ship[Supplier],Ship[Units Shipped]) |
| Calais | 30000 | =SUMIF([@Supplier],Ship[Supplier],Ship[Units Shipped]) |
| Robert | 2700 | =SUMIF([@Supplier],Ship[Supplier],Ship[Units Shipped]) |
| Simpson | 2300 | =SUMIF([@Supplier],Ship[Supplier],Ship[Units Shipped]) |

*Figure 8:Calculations in the SupSummary Table of the Technical Design. (To help the reader interpret column I, columns G and H show values although the cells actually contain formulas pointing to the raw data.)*

These table-style formulas appear cryptic until one learns how they work. In practice we never write the cell references manually, but instead click on cells or ranges as necessary, and allow Excel to generate the formula.

Notice that the cell formulas in each column are identical. This a powerful feature of Excel tables. The analyst need code a cell formula once in any cell of the column, and Excel automatically copies it to all other cells in that column of the table.

Regarding *accuracy*, coding is remarkably easy because references in cell formulas are generated automatically by Excel by clicking on the cell to be referenced, so the analyst need not attempt to write table-style formulas. Notably, the modest risk posed by copying formulas is eliminated entirely. As with the canonical design, the analyst must write simple pointer formulas to arrange the data; but of course this is minimally necessary for any design that is not data-driven.

The real significance of adding the abstraction of the Table is that the dimensionality and calculations of this model are entered once in the top row of the table and Excel automatically copies the formula down. Thus, *modification* is trivially easy. This is a significant prize. We illustrate with an example, where an additional supplier called Paulucci and an additional destination called Duluth are to be added. See Figure 9.

Shipping Costs and Sourcing Plan

| Row | Supplier | Destination | Shipping Cost/ Unit ($) | Units Shipped |
|---|---|---|---|---|
| 5 | Georgican | Chest | $ 30.80 | 1000 |
| 6 | Hickock | Chest | $ 36.80 | 1000 |
| 7 | India | Chest | $ 34.00 | 1000 |
| 8 | Johnson | Abbot | $ 42.00 | 1000 |
| 9 | Johnson | Bone | $ 41.60 | 1000 |
| 10 | Johnson | Chest | $ 45.60 | 1000 |
| 11 | Lincoln | Abbot | $ 33.15 | 1000 |
| 12 | Manister | Abbot | $ 32.00 | 1000 |
| 13 | Ocean | Abbot | $ 44.10 | 1000 |
| 14 | Ocean | Bone | $ 45.36 | 1000 |
| 15 | Calais | Bone | $ 35.00 | 1000 |
| 16 | Robert | Bone | $ 33.12 | 1000 |
| 17 | Simpson | Bone | $ 32.40 | 1000 |
| 22 | Paulucci | Abbot | 43 | |
| 23 | Paulucci | Bone | 40.75 | |
| 24 | Paulucci | Duluth | 35.5 | |

Supplier Table

| Row | Supplier | Supplier Capacity | Supplied |
|---|---|---|---|
| 5 | Georgican | 3,600 | 1000 |
| 6 | Hickock | 2,500 | 1000 |
| 7 | India | 9,000 | 1000 |
| 8 | Johnson | 3,000 | 3000 |
| 9 | Lincoln | 27,000 | 1000 |
| 10 | Manister | 6,000 | 1000 |
| 11 | Ocean | 3,000 | 2000 |
| 12 | Calais | 30,000 | 1000 |
| 13 | Robert | 2,700 | 1000 |
| 14 | Simpson | 2,300 | 1000 |
| 22 | Paulucci | 15000 | |

Destination Table

| Row | Destination | Required | Delivered |
|---|---|---|---|
| 5 | Abbot | 32,422 | 4000 |
| 6 | Bone | 21,233 | 5000 |
| 7 | Chest | 25,125 | 4000 |
| 22 | Duluth | 12555 | |

Cost Table

| Row | Total Sourcing Costs |
|---|---|
| 5 | $ 485,930 |

*Figure 9:Technical Design: New Data in rows 22:24*

Excel Tables have the property that new data can be pasted (or drag-and-dropped) beneath a table, and the table expands automatically. Excel automatically generates any necessary cell formulas in columns that contain formulas. Excel even adjusts the formatting of the data automatically!

The analyst need only to select cells B23:D25 and drag them to cells B18:D20, then perform an analogous drag-and-drop (or copy/paste or cut/paste) for the other two tables, and the modification task is completed. (We manually entered values for Units Shipped in new cells E18:E20.) All the formulas update, all the calculated values update, all the number formats are adjusted. Not a single formula needs to be touched!

**Shipping Costs and Sourcing Plan**

| Supplier | Destination | Shipping Cost/ Unit ($) | Units Shipped |
|---|---|---|---|
| Georgican | Chest | $ 30.80 | 1000 |
| Hickock | Chest | $ 36.80 | 1000 |
| India | Chest | $ 34.00 | 1000 |
| Johnson | Abbot | $ 42.00 | 1000 |
| Johnson | Bone | $ 41.60 | 1000 |
| Johnson | Chest | $ 45.60 | 1000 |
| Lincoln | Abbot | $ 33.15 | 1000 |
| Manister | Abbot | $ 32.00 | 1000 |
| Ocean | Abbot | $ 44.10 | 1000 |
| Ocean | Bone | $ 45.36 | 1000 |
| Calais | Bone | $ 35.00 | 1000 |
| Robert | Bone | $ 33.12 | 1000 |
| Simpson | Bone | $ 32.40 | 1000 |
| Paulucci | Abbot | $ 43.00 | 1000 |
| Paulucci | Bone | $ 40.75 | 1000 |
| Paulucci | Duluth | $ 35.50 | 1000 |

**Supplier Table**

| Supplier | Supplier Capacity | Supplied |
|---|---|---|
| Georgican | 3,600 | 1000 |
| Hickock | 2,500 | 1000 |
| India | 9,000 | 1000 |
| Johnson | 3,000 | 3000 |
| Lincoln | 27,000 | 1000 |
| Manister | 6,000 | 1000 |
| Ocean | 3,000 | 2000 |
| Calais | 30,000 | 1000 |
| Robert | 2,700 | 1000 |
| Simpson | 2,300 | 1000 |
| Paulucci | 15,000 | 3000 |

**Destination Table**

| Destination | Required | Delivered |
|---|---|---|
| Abbot | 32,422 | 5000 |
| Bone | 21,233 | 6000 |
| Chest | 25,125 | 4000 |
| Duluth | 12,555 | 1000 |

**Cost Table**

| Total Sourcing Costs |
|---|
| $ 605,180 |

*Figure 10:Technical Design: Modification complete without manual editing of cell formulas*

The technical approach is useful for *analysis*, but has the same caveat as with the data-driven approach in that it lacks the row-and-column visual guidance of the matrix. This approach seems to be easy for *transfer*. Indeed, the use of the model requires only the ability to move data so that no programming is required for any new situation!

A high degree of training and sophistication is required to implement the technical approach. The analyst must be familiar with Excel Tables, be able to express the model in a more abstract Tables-based structure, and recognize how to adapt new data to the format required by the Tables. Due to the ease of *modification*, it can likely be *transferred* to technically unsophisticated personnel.

## 4. Comparison of Designs and Recommendations

We summarize the three designs in terms of each of the quality criteria.

### 4.1 Accuracy -- Efficient and Correct Coding of the Model

The *Data-Driven* Design approach is fraught with the risk of errors due to the need to write multiple custom cell formulas by hand, especially as the number of suppliers and/or destinations grows. It does not require the use of pointer formulas to arrange the data, because the data is used as-is. The *Canonical Design* is the most intuitive approach and is relatively easy to build, with the primary effort being to organize the input data. It is necessary to write simple pointer formulas to arrange the data in the desired form. The *Technical Design* is extremely efficient and accurate. It is again necessary to write simple pointer formulas to arrange the data in the desired form.

### 4.2 Efficient and Accurate Model Modification in the Context of Operational Changes

The *Data-Driven Design* can accommodate new data easily as long as the structure of the inputs (valid supplier-destination combinations) does not change. However, such models are fragile, and any changes to supplier-destination combinations, new suppliers, or new destinations require a great deal of care and effort. The *Canonical Design* can accommodate new data – including new connections between existing suppliers and destinations – relatively easily. However, the process of adding new suppliers and/or destinations is somewhat cumbersome and risky. Because the *Technical Design* is agnostic as to the supplier-destination combinations, it can trivially accommodate anything, including new data, new supplier-destination combinations, new suppliers, and new destinations – all with no coding required.

### 4.3 Analysis and Generation of Business Insight

The matrix structure of the *Canonical Design* makes interactive "What-If" analysis easy and intuitive. Both the *Data-Driven Design* and *Technical Design* have a columnar structure that lacks the same visual intuition. In our experience this is not a problem for a sophisticated analyst. We hypothesize that one can use lookup functions to easily code a matrix-style report (ala the Canonical Design); this is a suitable topic for future research.

### 4.4 Transfer by the Spreadsheet Author for Use by Other Personnel

The major worry about transferring a model that follows a *Data-driven Design* is that any change whatsoever to the supplier-destination combinations will require manual programming, and may undermine the confidence of non-authors. In an environment where new suppliers and destinations are rarely introduced, the intuitive nature of the *Canonical Design* makes it easy to transfer to other personnel. From our experience, this representation of the problem in terms of supplier-destination pairs motivates user interaction with the model. On the other hand, in an environment where suppliers and/or destinations are frequently added, the *Technical Design* will be advantageous, as model users can easily be trained to update it themselves when structural changes take place, since one need merely move the data and need not touch any cell formulas.

## 5. Conclusions and Areas for Future Research

In this paper, we introduce the transportation model, which is used for a variety of practical operational problems and is often implemented in a spreadsheet. Using a simple but realistic illustrative example, we compare three different spreadsheet designs according to the criteria of accuracy, modification, analysis, and transfer, and we comment on the training and sophistication necessary to use each approach.

We believe that the flawed data-driven design has as its root cause a naïve approach to spreadsheet programming, where the author takes the structure of the data as a given, and starts to code directly without reflection about the implications of the resulting design. This perspective on data-driven design to spreadsheet development might be generalizable to other situations, and is an exciting area for further study.

We see a clear trade-off between the intuitive nature of the Canonical Design and the powerful flexibility that comes with the increased abstraction of our Technical Design. For relatively basic transportation problems such as our example problem, the choice between these two approaches depends largely on the sophistication of the model builders and end users and on the relative stability of the business environment. Each of these design approaches has its strengths and weaknesses.

Further research is required to understand the ease of analysis for the Technical Design, and in particular the possibility of coding a report that organizes the results into the intuitive matrix structure of the Canonical Design. We hypothesize that this can be done in a way that is amenable to *accuracy*, *modification* and *transfer*, but that remains to be demonstrated. This approach merits field testing or even empirical experiments; we will start that journey presently by teaching this approach to business students, which is likely to provide much insight.

The transportation model presented in this paper is a special case of a more general class of "network flow models" ([Ford and Fulkerson, 1962], [Edmonds and Karp, 1972], [Ahuja et al., 1989]) which have more complex "to/from" patterns and richer constraint sets. Network flow models have an even wider range of application than the transportation model. The insights in this paper for transportation models likely apply broadly to network flow models. Further research is needed to explore any limitations and required modifications or enhancements, particular in the context of additional conditions that are often referred to as "side constraints."

It has been hypothesized (Siersted 2015) that the existing methodologies for financial planning models (notably FAST) have applicability to any analytical spreadsheet model. The operations management model presented in this paper can serve as a basis for future research to test this hypothesis. This line of research has potential to help us better distinguish between principles that are specific to financial planning models and those that are more generally applicable.